\documentclass[12pt]{article}

\usepackage{graphicx}
\usepackage{amsmath,amssymb}
\usepackage{amsthm,amscd,mathrsfs}

\newcommand{\be}{\begin{eqnarray}} 
\def\ne{\nonumber\end{eqnarray}}
\newcommand{\ee}{\end{eqnarray}}
\newcommand{\nn}{\nonumber}
\def\l{\left}
\def\r{\right}
\def\hf{{1\over2}}
\def\d{\partial}
\def\cd{\nabla}

\newcommand{\gsim}{\mathrel{\mathop{\kern-0pt >}\limits_{\sim}}}
\newcommand{\lsim}{\mathrel{\mathop{\kern-3pt <}\limits_{\sim}}}
\def\sitarel#1#2{\mathrel{\mathop{\kern-0pt #1}\limits_{#2}}}
\def\inbar{\,\vrule height1.5ex width.4pt depth0pt}
\def\IC{\relax\hbox{$\inbar\kern-.3em{\rm C}$}}
\def\IH{\relax{\rm I\kern-.18em H}}
\def\IP{\relax{\rm I\kern-.18em P}}
\def\IQ{\relax\hbox{$\inbar\kern-.3em{\rm Q}$}}
\def\IR{{\bf R}}
\font\cmss=cmss10 \font\cmsss=cmss10 at 7pt
\def\IZ{\relax\ifmmode\mathchoice
{\hbox{\cmss Z\kern-.4em Z}}{\hbox{\cmss Z\kern-.4em Z}}
{\lower.9pt\hbox{\cmsss Z\kern-.4em Z}}
{\lower1.2pt\hbox{\cmsss Z\kern-.4em Z}}\else{\cmss Z\kern-.4em Z}\fi}
\def\1{{\bf 1}}

\def\G{\Gamma}
\def\g{\gamma}
\def\s{\sigma}
\def\dl{\delta}
\def\La{\Lambda}
\def\la{\lambda}
\def\e{\epsilon}

\def\a{\alpha}

\def\da{\dot{\alpha}}
\def\db{\dot{\beta}}
\def\om{\omega}
\def\N{{\cal N}}
\def\L{{\cal L}}
\def\bz{{\bar{z}}}
\def\bzz{{\bar{z}z}}
\def\tr#1{{\rm tr}{\left[{#1}\right]}}
\def\t#1{\tilde{#1}}

\begin{document}

\vspace*{-1.2cm}
\hfill{UT-12-17}\\

\vskip 1cm
\centerline{\Large\bf 
5D SYM on 3D Sphere and 2D YM
}

\vskip .5in

\centerline{\sc Teruhiko Kawano ~and~ Nariaki Matsumiya}

\vskip .1in
\centerline{\it Department of Physics, University of Tokyo, 
Hongo, Tokyo 113-0033, Japan}

\vskip .5in
{\small 
It is shown by using localization 
that in five-dimensional $\N=1$ supersymmetric Yang-Mills theory 
on ${S}^3$, correlation functions in a sector 
are identical to correlation functions in 
two-dimensional bosonic Yang-Mills theory. 
}
\vskip .3in
\setcounter{footnote}{0}
\section{Introduction}

It has been observed that a correlation function in a gauge theory 
gives the same result as the one in a matrix theory. 
The recent typical examples in our mind are four-dimensional 
$\N=4$ supersymmetric Yang-Mills theory on $S^4$ \cite{Pestun} and 
the ABJM model on $S^3$ \cite{Kapustin}, 
where the localization method was used to calculate 
the correlation functions exactly. 
A similar application of the localization method to five-dimensional 
supersymmetric Yang-Mills theory on $S^5$ has been attempted in 
\cite{Terashima}. 

Contrary to an application of the localization method to an $n$-dimensional 
supersymmetric gauge theory on an $n$-dimensional compact space, which 
has been seen to give rise to a matrix model, 
we will in this paper study five-dimensional $\N=1$ 
supersymmetric Yang-Mills theory put on Euclidean $\IR^2\times{S}^3$ 
by using the localization method. 

To this end, one needs to pick up a supersymmetry as 
the BRST symmetry, and a correlation function of the BRST invariant 
operators can be calculated by using the localization method. 
In this paper, it will be seen that a correlation function of 
the BRST invariant operators in the supersymmetric Yang-Mills theory 
yields the one in two-dimensional bosonic Yang-Mills theory.

\section{Five-Dimensional Super Yang-Mills Theory on $\IR^2\times{S}^3$}

A vector supermultiplet in the five-dimensional $\N=1$ Yang-Mills theory 
consists of a gauge field $v_M$, a real scalar field $\s$, an auxiliary field 
$D^{\da}{}_{\db}$, and a spinor field $\Psi^{\da}$, 
where the indices $\da$, $\db$ label the components of 
the fundamental representation ${\bf 2}$ of $SU(2)$ $R$-symmetry. 
The spinor field obeys the symplectic Majorana condition
\be
(\Psi^{\dot{\beta}})^T C_5\epsilon_{\dot{\beta}\dot{\alpha}}
=\l({\Psi}{}_{\dot{\alpha}}\r)^{\dag}
\equiv\bar{\Psi}{}_{\dot{\alpha}},
\ne
where $T$ denotes the transpose, and $\e_{\da\db}$ is the invariant 
tensor of the $SU(2)$ R-symmetry. 
The auxiliary field $D^{\da}{}_{\db}$ is 
anti-Hermitian and in the adjoint representation of $SU(2)$ R-symmetry;
\be
D^{\da}{}_{\db}=-(D^{\db}{}_{\da})^\dag,
\qquad
D^{\da}{}_{\dot\g}\,\e^{\dot\g\db}=D^{\db}{}_{\dot\g}\,\e^{\dot\g\da},
\qquad
D^{\da}{}_{\da}=0. 
\ne
Our notations for the charge conjugation 
matrix $C_5$ and the gamma matrices $\G^M$ are explained 
in \ref{notations}. 
We assume that the gauge group $G$ is a simple Lie group. 
All the fields are in the adjoint representation of the gauge group $G$, 
and are denoted in the matrix notation as 
\be
\Phi=\Phi^AT^A
\ne
with the normalization $\tr{T^AT^B}=\dl^{AB}$. 

On a flat Euclidean space $\IR^5$, the Lagrangian $\L_V$ is given 
by 
\begin{equation}
\tr{
\frac{1}{4}v_{MN}v^{MN}-\frac{1}{2}D_M \s D^M \s 
-i\bar{\Psi}{}_{\dot{\alpha}}\G^M D_M\Psi^{\dot{\alpha}}
+g\bar{\Psi}{}_{\dot{\alpha}}\l[\s,\,\Psi^{\dot{\alpha}}\r]
+\frac{1}{4}D^{\dot{\alpha}}{}_{\dot{\beta}}D^{\dot{\beta}}{}_{\dot{\alpha}}, 
}
\label{flatLv}
\end{equation}
where $v_{MN}$ is the field strength 
\be
v_{MN}=\d_Mv_N-\d_Nv_M+ig\l[v_M,\,v_N\r],
\ne
of the gauge field $v_M$, and the covariant derivatives 
$D_M\Phi$ is given by
\be
D_M\Phi=\d_M\Phi+ig\l[v_M,\,\Phi\r].
\ne

The Lagrangian $\L_V$ is left invariant under a supersymmetry transformation
\be
&&\delta^{(0)} v_M=-i\bar{\Sigma}_{\dot{\alpha}}\G_M\Psi^{\dot{\alpha}},
\qquad
\delta^{(0)} \sigma=i\bar{\Sigma}_{\dot{\alpha}}\Psi^{\dot{\alpha}}, 
\nn\\
&&\delta^{(0)}\Psi^{\dot{\alpha}}
=-\hf\l(\hf{}v_{MN}\G^{MN}\Sigma^{\dot{\alpha}}
+\G^MD_M\s\Sigma^{\dot{\alpha}}+D^{\da}{}_{\db}\Sigma^{\db}\r),
\label{flatSUSY}\\
&&\delta^{(0)} D^{\dot{\alpha}}{}_{\dot{\beta}}
=i\l[D_M\bar{\Psi}{}_{\dot{\beta}}\G^M\Sigma^{\dot{\alpha}}
+\bar{\Sigma}_{\dot{\beta}}\G^M D_M\Psi^{\dot{\alpha}}
+ig\l(\l[\s,\,\bar{\Psi}{}_{\dot{\beta}}\r]\Sigma^{\dot{\alpha}}
+\bar{\Sigma}_{\dot{\beta}}\l[\s,\,\Psi^{\dot{\alpha}}\r]\r)\r], 
\ne
where the transformation parameter $\Sigma^{\da}$ is also a symplectic 
Majorana spinor;
\be
\bar{\Sigma}{}_{\dot{\alpha}}=
(\Sigma^{\dot{\beta}})^T C_5\epsilon_{\dot{\beta}\dot{\alpha}}.
\ne

When the system is put on $\IR^2\times{S}^3$, it is convenient to 
give the gauge field $v^M$ and the spinor $\Psi^{\da}$ in terms of 
three-dimensional tensors and spinors as
\be
&&v^{m}\quad (m=1,2,3),
\qquad
v_z=\hf\l(v_4-iv_5\r),
\qquad
v_{\bz}=\hf\l(v_4+iv_5\r),
\nn\\
&&\Psi^{\da=1}=\lambda\otimes\chi_++
\psi\otimes\chi_-,
\qquad
\Psi^{\dot{\alpha}=2}=C^{-1}_3 \psi^{*}\otimes\chi_{+}
+C^{-1}_3 \lambda^{*}\otimes\chi_{-},
\nn\\
&&D=D^{1}{}_{1}+2\,v_{\bzz},
\qquad
F=\hf\,D^{1}{}_{2}, 
\qquad
\bar{F}=\hf\,D^{2}{}_{1}, 
\ne
where $*$ denotes the complex conjugation, and 
the complex coordinates $z,\bz$ for $\IR^2$ were introduced 
by $z=x^4+ix^5$, $\bz=x^4-ix^5$. The two-dimensional spinors 
\be
\chi_{\pm}={1\over\sqrt{2}}\left(\begin{array}{c}
1\\{\pm{i}}\end{array}\right),
\ne
are the eigenvectors of $i\G^4\G^5$; 
$
i\G^4\G^5\chi_{\pm}=\pm\chi_{\pm}.
$

For the supersymmetry transformation (\ref{flatSUSY}), 
in going onto the $\IR^2\times{S}^3$, we will pick up 
one of the Killing spinors $\e$ on $S^3$ obeying that
\be
\cd_m\e={i\over2}\g_m\e,
\ne
and set 
\be
\Sigma^{\da=1}=\e\otimes\chi_+,
\qquad
\Sigma^{\da=2}=C_3^{-1}\e^{*}\otimes\chi_-.
\ne

On the $\IR^2\times{S}^3$, 
the supersymmetry transformation (\ref{flatSUSY})
no longer yields a closed algebra. Note here that the covariant derivative 
$D_m\la$ contains the spin connection $\om_m$ of the unit round $S^3$ as
\be
D_m\la=\d_m\la+{1\over4}\om_m^{ab}\g^{ab}\la,
\ne
where $a,b=1,2,3$ denote the tangent indices.  
In order to obtain a closed algebra, 
we will modify the transformation law of $D^{\da}{}_{\db}$ by adding 
\be
\dl^{\prime}_{\epsilon} D 
=-\hf\l(\bar{\e}\la -\bar{\la}\e\r),
\qquad
\dl^{\prime}_{\epsilon} F =\hf\e^T C_3 \psi,
\ne
to them, respectively. 

One then finds that the modified transformation 
$\dl_\e=\dl^{(0)}_\e+\dl'{}_\e$, 
\be
&&\dl_\e v_m=-i\l[\bar{\e}\g_m\la-\bar{\la}\g_m\e\r],
\qquad
\dl_\e v_z=-\bar{\e}\psi, 
\qquad
\dl_\e\s =i\l[\bar{\e}\la-\bar{\la}\e\r],
\nn\\
&&\dl_\e\la=-\hf\l[\hf\,v_{mn}\g^{mn}
+\g^mD_m\sigma+D\r]\e,
\nn\\
&&\dl_\e\psi=-\l[-iv_{mz}\g^m\e+iD_z\s\e+F\,C^{-1}_3\e^{*}\r],
\label{SUSY}\\
&&\dl_\e D=i\bigg[D_m \bar{\la}\g^m\e+\bar{\e}\g^mD_m\la
+ig\l(\l[\s,\,\bar{\lambda}\r]\epsilon+\bar{\e}\l[\s,\,\la\r]\r)
+{i\over2}\l(\bar{\e}\la -\bar{\la}\e\r)
\bigg],
\nn\\
&&\delta_\e F =i\l[-\epsilon^T C_3\g^m D_m\psi^a
+2i\epsilon^TC_3D_z\lambda^a+ig\epsilon^TC_3\l[\s,\,\psi\r]
-{i\over2}\e^T C_3 \psi
\r],
\ne
yields the closed algebra
\be
&&\l[\dl_{\eta},\,\dl_{\e}\r]
v_m 
= -i\l(\xi^n\cd_n v_m-D_m\om\r)-\l(\bar{\e}\g_{mn}\eta
-\bar{\eta}\g_{mn}\e\r)v^{n}, 
\nn\\
&&\l[\dl_{\eta},\,\dl_{\e}\r]v_z 
= -i\l(\xi^n \cd_n v_z -D_z\om\r), 
\qquad
\l[\dl_{\eta},\,\dl_{\e}\r]\sigma 
= -i\l(\xi^n \nabla_n \sigma +ig\l[\om,\, \s\r]\r), 
\nn\\
&&\l[\dl_{\eta},\,\dl_{\e}\r]\la 
= -i\l(\xi^n\cd_n\la+ig\l[\om,\,\la\r]\r)
-\l(\bar{\e}\eta-\bar{\eta}\e\r)\la
-{1\over4}\l(\bar{\e}\g_{mn}\eta-\bar{\eta}\g_{mn}\e\r)\g^{mn}\la, 
\nn\\
&&\l[\dl_{\eta},\,\dl_{\e}\r]\psi 
= -i\l(\xi^n\cd_n\psi+ig\l[\om,\,\psi\r]\r)
-\l(\bar{\e}\eta-\bar{\eta}\e\r)\psi
-{1\over4}\l(\bar{\e}\g_{mn}\eta-\bar{\eta}\g_{mn}\e\r)\g^{mn}\psi, 
\nn\\
&&\l[\dl_{\eta},\,\dl_{\e}\r]D 
= -i\l(\xi^n\cd_n D+ig\l[\om,\,D\r]\r), 
\nn\\
&&
\l[\dl_{\eta},\,\dl_{\e}\r]F
= -i\l(\xi^n\cd_n F+ig\l[\om,\,F\r]\r)-2\l(\bar{\e}\eta-\bar{\eta}\e\r)F, 
\ne
with the transformation parameters
\be
\xi^m=\bar{\e}\g^m\eta-\bar{\eta}\g^m\e,
\qquad
\om=\xi^n v_n +\l(\bar{\e}\eta-\bar{\eta}\e\r)\s, 
\ne
where the covariant derivative $\cd_mv_n$ includes the Levi-Civita 
connection $\G^k{}_{mn}$ as
\be
\cd_mv_n=\d_mv_n-\G^k{}_{mn}v_k.
\ne

On the $\IR^2\times{S}^3$, since the three-dimensional sphere $S^3$ is 
a curved space, one needs to replace all derivatives on spinors in 
the Lagrangian $\L_V$ by the covariant derivative with the spin connection 
$\om_m$. However, it isn't enough to be invariant under the modified 
supersymmetry transformation (\ref{SUSY}). In fact, to the original 
Lagrangian $\L_V$,
\be
&&{\rm tr}\bigg[
\frac{1}{4}(v_{mn})^2+2\l|v_{mz}\r|^2
-\frac{1}{2}(D_m\sigma)^2-2D_z\sigma D_{\bar{z}}\sigma 
+\hf{}D^2-2v_{\bzz}\,D
+2\bar{F}F
\nn\\
&&\hskip0.7cm
-2i\bar{\la}\g^mD_m\la+2i\bar{\psi}\g^mD_m\psi
+4\bar{\psi}\,D_z\la+4D_{\bz}\bar\la\,\psi
+2g\l(\bar{\la}\l[\s,\,\la\r]+\bar{\psi}\l[\s,\,\psi\r]\r)
\bigg]
\ne
one needs to add 
the two terms
\be
\L'{}_V
=-\tr{\bar{\psi}\psi+\bar{\la}\la+\s\s+i\s\l(D-4v_{\bzz}\r)},
\quad
{\L}_{CS}=-\e^{mnl}\tr{v_m \d_n v_l +i\frac{g}{3}v_m\l[v_n,\,v_l\r]},
\ne
to obtain the supersymmetric total Lagrangian 
\be
\L=\L_V+\L'{}_V+\hf\L_{CS}.
\ne
One can then verify that $\dl_\e\L=0$.

\section{Localization}

In this section, we will calculate the partition function of 
the five-dimensional supersymmetric Yang-Mills theory on the 
$\IR^2\times{S}^3$ by using the localization method. 
In order to make the path integral well-defined, the bosonic fields 
$\s$, $D$, $F$, and $\bar{F}$ need to be analytically continued. 
Therefore, we will regard the scalar field $\s$ as taking pure imaginary 
values, and the auxiliary field $D$ as a real field. 
Further, $\bar{F}=F^*$.

To carry out the localization method, we will define the BRST transformation 
by setting $\bar\e$ to zero in the supersymmetric transformation (\ref{SUSY}) 
and by replacing the Grassmann odd parameter $\e$ by a Grassmann even one.
It yields 
\be
&&\dl_Q v_m=-i\bar{\la}\g_m\e,
\qquad
\dl_Q v_z=0, 
\quad
\dl_Q v_{\bz}=\bar\psi\e, 
\qquad
\dl_Q\s =i\bar{\la}\e,
\nn\\
&&\dl_Q\la=-\hf\l[\hf\,v_{mn}\g^{mn}+\g^mD_m\sigma+D\r]\e,
\qquad
\dl_Q\bar\la=0,
\nn\\
&&\dl_Q\psi=i\l[v_{mz}\g^m-D_z\s\r]\e,
\qquad
\dl_Q\bar\psi=\bar{F}\,\e^TC_3,
\label{BRST}\\
&&\dl_Q D=-i\bigg[D_m \bar{\la}\g^m\e
+ig\l[\s,\,\bar{\lambda}\r]\e-{i\over2}\bar{\la}\e
\bigg],
\nn\\
&&\delta_Q F =i\e^TC_3\l[-\g^m D_m\psi^a
+2iD_z\lambda^a+ig\l[\s,\,\psi\r]
-{i\over2}\psi\r],
\quad
\delta_Q \bar{F} =0,
\ne
which is in fact nilpotent; $\dl_Q^2=0$, as it should be. 
Using the BRST transformation (\ref{BRST}), we will modify the Lagrangian $\L$ 
into $\L+{t}\L_Q$ with a parameter $t$, where 
\be
\L_Q=\dl_Q\tr{
\l(\dl_Q\la\r)^\dag\la+\l(\dl_Q\psi\r)^\dag\psi+
\bar\psi\l(\dl_Q\bar\psi\r)^\dag}.
\ne
The bosonic part of the extra Lagrangian $\L_Q$ gives 
\be
&&\L_Q^{(B)}
=\hf{\rm tr}\bigg[
{1\over4}\l(v_{mn}\r)^2
+2\l|v_{mz}\r|^2
-\hf{D}_m\s{D}_m\s
-2D_z\s D_{\bz}\s
+\hf\,D^2
\nn\\
&&\hskip4.5cm
+2\l|F\r|^2
+2k_m\l(v_{mz}D_{\bz}\s-v_{m\bz}D_z\s+i\e_{mnk}v_{n\bz}v_{kz}\r)
\bigg]
\ne
where the Killing vector $k_m$ was defined by
\be
k_m=\bar\e\g_m\e
\ne
with the normalization $\l(\bar\e\e\r)=1$.
On the other hand, the fermionic part of $\L_Q$ gives 
\be
\L_Q^{(F)}=
i{\rm tr}\bigg[
-\bar\la\g^m{D}_m\la-{i\over2}\bar\la\la-ig\bar\la\l[\s,\,\la\r]
+\bar\psi\g^m{D}_m\psi-{i\over2}\bar\psi\psi-ig\bar\psi\l[\s,\,\psi\r]
\nn\\
-ik_m\bar\psi\g^m\psi+2i\bar\la{D}_{\bz}\psi
-i\bar\psi{D}_z\la+ik_m\bar\psi\g^mD_z\la
\bigg]. 
\ne

In the large $t$ limit, $t\to\infty$, the fixed point, 
which is a solution to 
\be
\l[\hf\,v_{mn}\g^{mn}+\g^mD_m\sigma
+D\r]\e=0,
\qquad
\l[v_{mz}\g^m-D_z\s\r]\e=0,
\qquad
F=0,
\ne
gives
the dominant contribution to the partition function. 
In fact, the fixed point is given by
\be
&&v_m=0, 
\quad
D=0,
\quad
F=0,
\quad
v_z=v_z(z,\bz), 
\quad
\s=\s(z,\bz), 
\quad
D_z\s=0.
\label{fixedpt}
\ee

Substituting the background (\ref{fixedpt}) into the original Lagrangian $\L$, 
one finds that the additional Lagrangian $\L'{}_V$ only contributes and yields 
\be
\L_{YM}=\tr{-\s\s+4i\s\,v_{\bzz}},
\label{YM}
\ee
which is the action of the two-dimensional Yang-Mills theory after eliminating 
the scalar field $\s$.

Around the fixed points, one needs to evaluate the path integral 
over the quantum fluctuations. Since the bosonic fields $\s$, $v_z$, 
and $v_{\bz}$ have a non-trivial background as the fixed point, 
we will expand the fields as
\be
\s=\s(z,\bz)+{1\over\sqrt{t}}\t{\s}(x^m,z,\bz), 
\qquad
v_z=v_z(z,\bz)+{1\over\sqrt{t}}\t{v}_z(x^m,z,\bz), 
\ne
while the other fields are rescaled as 
$\Phi\to(1/\sqrt{t})\tilde\Phi$, as in \cite{Kapustin}.

One also needs the gauge-fixing procedure for the evaluation of 
the path integral. We will follow \cite{Kapustin} and add 
to $\L_Q$ the gauge-fixing term and the ghost term 
\be
\tr{\bar{c}\cd_m{D}^mc+B\cd^mv_m}. 
\ne

There remains the residual gauge symmetry, under which 
\be
\s\to\s+ig\l[\om(z,\bz),\s\r], 
\qquad
v_z\to{v}_z-D_z\om(z,\bz),
\label{residualgauge}
\ee
where the gauge transformation parameter $\om$ is constant on the $S^3$. 
Following \cite{Lecture,qYM}, 
one can make use of the residual symmetry (\ref{residualgauge}) 
and $D_z\s=0$ in (\ref{fixedpt}) to put 
the background $\s(z,\bz)$, $v_z(z,\bz)$ in the Cartan subalgebra 
of the Lie algebra of $G$ such that 
\be
\s(z,\bz)=\sum_{i=1}^{r}\s_i\,{}H_i, 
\qquad
v_z(z,\bz)=\sum_{i=1}^{r}v^i{}_z(z,\bz){}H_i, 
\label{Cartan}
\ee
where $H_i$ ($i=1,\cdots,r$) are the generators of 
the Cartan subalgebra of rank $r$, and $\s_i$ ($i=1,\cdots,r$) are 
constant with respect to $z,\bz$. 

Therefore, for the residual gauge symmetry (\ref{residualgauge}), 
we will follow the same BRST quantization procedure 
as for the two-dimensional Yang-Mills theory 
in \cite{Lecture,qYM}. The path-integral measure of the scalar field 
$\s(z,\bz)$ thus results in the finite-dimensional integral over 
$\s_i$ ($i=1,\cdots,r$) and the determinant of the Fadeev-Popov ghosts. 

One can thus see that the localization procedure has so far given 
the same Lagrangian (\ref{YM}), the same fixed points (\ref{Cartan}), 
and the same BRST gauge fixing procedure as for the Yang-Mills 
theory, - the exactly same results as in \cite{Lecture,qYM}, but, except for 
one point. In the two-dimensional Yang-Mills theory, 
for the two-dimensional gauge fields $v_z$, $v_{\bz}$, 
the root part 
\be
\sum_{\a\in\La}v^{\a}{}_z(z,\bz)\,E_\a,
\qquad
\sum_{\a\in\La}v^{\a}{}_{\bz}(z,\bz)\,E_\a,
\label{rootzeromode}
\ee
where $\La$ is the set of all the root of the Lie algebra of $G$, 
and the root generators $E_{\a}$ satisfy the algebra
\be
\l[H_i,\,E_{\a}\r]=\a_i\,{E}_\a, 
\qquad
\l[E_\a,\,E_{-\a}\r]=\sum_{i=1}^{r}\a_i\,{H}_i\equiv\a\cdot{H},
\ne
show up in the Lagrangian (\ref{YM}) as
\be
4g\l(\a\cdot\s\r)\l|v^{\a}{}_{z}\r|^2,
\label{discrepancy}
\ee
and yield the contributions to the partition function. 

However, in our case, it no longer gives any contributions in the large 
$t$ limit, $t\to\infty$. 

One then proceeds to the evaluation of the one-loop determinants 
from the Lagrangian $\L_Q$, which also contains the root part 
in (\ref{rootzeromode}) as the zero modes of $v_z$, $v_{\bz}$ 
upon expanding them in terms of the harmonics on $S^3$. 
To this end, we will follow the same procedure as in \cite{Kapustin,Hosomichi}, 
- expanding all the fields in terms of the harmonics on $S^3$ and 
performing the Gaussian integration over them. 

Up to an overall irrelevant normalization constant, 
the tedious calculation shows the exact cancellation between the bosonic 
degrees of freedom and fermionic ones, but, except for one pair. 
The zero modes of the scalar harmonics from $v_z$, $v_{\bz}$ do not 
cancel out the contribution of 
one of the low-lying modes of the spinor harmonics from $\la$, $\psi$ 
to yield the same contribution as the discrepancy, which would come from 
(\ref{discrepancy}) in the two-dimensional Yang-Mills theory;
\be
\int\l[d\Phi\r]e^{-\int\,d^5x\l(\L+t\L_Q\r)}
\quad\longrightarrow\quad
\int\prod_{\a\in\La}\l[dv^\a{}_{z}dv^{\a}_{\bz}\r]
e^{-\int{dzd\bz}\,4g\l(\a\cdot\s\r)\l|v^\a{}_z\r|^2}.
\ne
The partition function in the supersymmetric Yang-Mills theory 
on $\IR^2\times{S}^3$ thus exactly reduces into the one 
in the bosonic Yang-Mills theory on $\IR^2$, via the localization method.

\section{Discussions}

In the previous sections, we have seen that 
the five-dimensional $\N=1$ supersymmetric Yang-Mills theory 
can be put on the product space $\IR^2\times{S}^3$ with 
supersymmetry kept. Upon the localization, without losing any 
degrees of freedom, it reduces into the two-dimensional 
Yang-Mills theory; namely, a correlation function in the BRST 
invariant sector of the five-dimensional supersymmetric theory 
is identical to a correlation function in the two-dimensional 
bosonic theory. 

A conceivable extension of this work would be inclusion of 
hypermultiplets into the five-dimensional theory \cite{FKM}. 
It would be interesting to see what would happen 
in the two-dimensional theory and to understand the relation to 
the proposal in \cite{Rastelli}.

\vskip 2cm
\centerline{\bf Acknowledgement}

\medskip
The authors would like to thank Yasutaka Fukuda and Satoshi Yamaguchi 
for collaborations at the early stage of this work. 
We are grateful to Yuji Tachikawa for helpful discussions 
and for a careful reading of the manuscript. 
We are also grateful to Kazuo Hosomichi for helpful discussions 
and for giving us crystal clear lectures about the papers 
\cite{Kapustin,Hosomichi}, 
where we have learnt all the techniques we needed for this work. 
The work of T.~K. was supported in part 
by a Grant-in-Aid \#23540286 from the MEXT of Japan.

\appendix
\noindent
\def\thesection{Appendix\,{}}
\section{}
\label{notations}

The five-dimensional gamma matrices $\G^M$ ($M=1,\cdots,5$) 
satisfy  
\be
\l\{\G^M,\,\G^N\r\}=2\delta^{MN}, 
\ne
and they are given in terms of the three-dimensional gamma matrices 
$\g^m=\s_m$ ($m=1,2,3$) as
\be
\G^m=\g^m\otimes \s_2, 
\qquad
\G^4=\1\otimes\s_1,
\qquad
\G^5=\1\otimes\s_3,
\ne
where $\s_{1,2,3}$ are the Pauli matrices. 

The five-dimensional charge conjugation matrix $C_5$ satisfies 
\be
\l(\G^M\r)^T=C_5\, \G^M\, C_5^{-1},
\qquad
\l(C_5\r)^T=-C_5,
\ne
where $T$ denotes the transpose of the matrix, and it 
may be given in terms of the three-dimensional charge conjugate matrix 
$C_3=i\s_2$ as 
\be
C_5=C_3\otimes\1.
\ne

\clearpage

\end{document}